\documentstyle[12pt,aps,preprint]{revtex} 
\begin{document} 
\tightenlines
\title{Charging kinetics of dust particles with a variable mass}
\author{\large{S.A.Trigger \vskip 3mm }
\normalsize Institute for High Temperatures RAS, Izhorskaya 13/19, 127412
  Moscow, Russia}            
\maketitle
\vskip 1cm
\begin{list}{\underline{\bf{Keywords}}: \,\,\,\, Kinetics, Dusty Plasmas, Charging, Mass  Variable.\hfill\hfill\hfill}\item \end{list} 
\vspace{1.5cm} 

\begin{list}{{\bf Abstarct}} \item \end{list} 
A kinetic equation for dust particles with a new 
kinetic variable - the mass of grains - is suggested
for the description of processes with changing mass 
in dusty plasmas and neutral systems.\\         


\begin{list}{\bf I. Introduction\hfill\hfill\hfill}\item \end{list}  
\vskip 3mm
   The kinetic theory of dusty plasmas, which takes into 
account  specific processes of charging, has been considered 
in many papers, but usually not from first principles.
In [1] the dust charge was introduced as a new dynamic 
variable for the kinetic equation, in [2,3] collision 
integrals for dusty plasmas with charging have been 
formulated and used for several applications. The form 
of the charging collision integrals suggested in [2,3] 
has been recently rigorously justified  in [4,5], where 
also the stationary velocity and charge distributions 
for dusty plasmas were established.
In this report we consider the generalized kinetic equation 
in which a new dynamic variable - the mass of dust particles 
- is introduced in parallel with the charge variable. 
We will show that for models of dusty plasmas with absorption 
of ions and electrons the distribution function of grains and 
the average kinetic energy  are determined not only by 
momentum transfer from  light plasma particles to dust 
particles, but essentially (on the same time scale) also by 
mass transfer. 
This statement also agrees with the preliminary results of MD 
simulations of the heating of dust particles in plasma [6].
   A simplified form of the obtained kinetic equation (the 
nonstationary variant of the Fokker-Planck equation with 
changing mass) is also found and the simplest concrete 
applications are considered.
For some more complicated situations with surface chemical 
reactions between electrons and ions absorbed by dust 
particles, when atoms appear and can return from the dust to 
plasma, the processes of mass transfer can also be essential.
   It is necessary to emphasize that the formulated equation 
can be important for different applications not only for 
plasmas, but for other systems with mesoscopic particles, 
where processes with mass transfer take place. \\

\begin{list}{\bf II. Kinetic equation for dust particles with mass and
 charge variables\hfill\hfill\hfill}\item \end{list} 
\vskip 3mm
   Let us introduce the generalized kinetic equation in which 
a new dynamic variable $M$ - the mass of dust particles is 
included in parallel with the charge $Q$:

\begin{equation}
 \frac{df_D(t)}{dt} = J_D({\vec  p},{\vec r},t,Q,M) + J_D^c,
\end{equation}
where the collision integral $J_D^c$ describes all collision
processes without change of number of small particles (e.g.\
electrons and ions in plasmas) and without change of mass and
charge of grains. The collision integral $J_D$ describes the 
absorption of mass and charge by dust particles. For the 
simple model of absorption of electrons and ions with masses 
$m_{\alpha}$ and charges $e_{\alpha}$ by a grain with charge $Q$ we can write:

\begin{eqnarray}
J_D=&&\sum_{\alpha=(e,i)} \int d{\vec p} 
f_{\alpha}({\vec p},{\vec r},t)
 \lgroup \omega_{\alpha}({\vec p},{\vec P}-{\vec p},Q-e_{\alpha},M-m_{\alpha})
 f_D({\vec P}-{\vec p},{\vec r},t,Q-e_{\alpha},M-m_{\alpha}) \nonumber \\
&&  -\omega_{\alpha}({\vec p},{\vec P},Q,M)
 f_D({\vec P},{\vec r},t,Q,M)  \rgroup.
\end{eqnarray}
where $\omega_{\alpha}({\vec p},{\vec P},Q,M)=\upsilon
 \sigma_{\alpha}(\upsilon,Q,M)$
 is the probability density of absorption of an 
electron or ion with momentum $\vec p$ and charge $e_{\alpha}$ by a grain 
with momentum $\vec P$ and charge $Q$.
This collision integral implies that the processes of mass
transfer from grains back to the plasma are absent. If there
are such type of processes the more complicated equations can 
be written to take into account the mass balance in plasmas
correctly. The simplest approximation for the cross-section 
of absorption $\sigma_{\alpha}$ can be chosen in the usual form 
(see e.g. [3]).
In general the cross-section can be some function of the mass 
of grains (for example due to the dependence of the grain's 
radius on mass). In general the necessity to include the mass 
as a new kinetic variable depends, naturally, on the time 
scale, under consideration.
To simplify the problem we have to expand the collision 
integrals using the small parameters 
 $\frac{m_{\alpha}}{M}$ and
 $\frac{e_{\alpha}}{Q}$. We also 
suggest here that $P \gg p$.
Then we can find the generalized Fokker-Planck equation for 
grains, which will be nonstationary in our case, due in 
particular to mass absorption. In this paper we will realize 
this expansion for a neutral system: neutral grains 
in system of small neutral particles. Simplification of
Eq.1 and Eq.2 for plasmas is similar and will be published
separately.\\

\begin{list}{\bf III. Kinetics of neutral grains due to mass absorption collisions}\item \end{list} 
\vskip 3mm
   For neutral homogeneous systems we can rewrite Eq.2
in the form

\begin{eqnarray}
J_D(P,M,t)= &&\int d{\vec p} f_n({\vec p},t) \lgroup
  w ({\vec p},{\vec P}-{\vec p},M-m)
 f_D({\vec P}-{\vec p},M-m,t) \nonumber \\ &&
  - w( p,P,M) f_D( P,M,t)  \rgroup.
\end{eqnarray}
Here $w( p,P,M)=\sigma(M)\vert {{\vec P} \over M} -{{\vec p}
\over m} \vert$. 

After expansion up to first order in $\frac{m_{\alpha}}{M}$ and up to
second order in $\frac{p}{P}$ we find the kinetic equation for
grains:
\begin{eqnarray}
\frac{\partial f_D}{\partial t} = && \frac{j_{\epsilon} \sigma(M)}
{3} \Delta_P f_D(P,M,t)+\frac{j_0 m \sigma(M)}
{3M} \frac{\partial}{\partial P_{\alpha}}  
[P_{\alpha} f_D(P,M,t) ]  \nonumber \\
&& - j_0 m \frac{\partial}{\partial M}
[\sigma(M) f_D(P,M,t) ]\,\, ,
\end{eqnarray}
where:
\begin{equation}
j_0 = \int d{\vec p} \frac{p}{m} f_n(p,t), \,\,\,\,\,\,\,
j_{\epsilon} = \int d{\vec p} \frac{p^3}{2m} f_n(p,t). 
\end{equation}
 
Below we will suggest that $f_n(p,t)=f_n(p)$ is stationary Maxwell 
distribution for small particles with temperature $T_0$ and
density $n_0$. Then we find:
\begin{equation}
j_0 = \frac{4 n_0 T_0^{1/2}}{\sqrt{2 \pi m}},\,\,\,\,\,\,\,\,\,\,
j_{\epsilon} = 2mT_0 j_0 .
\end{equation} 

Let us consider some average functions: density $n_D$, 
mass of grain $<M>\,\,\,$ and \\
$U_{ab}(\lambda)=<\frac{p^a}{\lambda M^b}>$,
 where we define the averaging
as
\begin{equation}
n_D = \frac{1}{M_0} \int d {\vec P} d M f_D (P,M,t), \,\,\,\,\,\,\,\,
<A> = \frac{1}{n_D M_0} \int d {\vec P} d M f_D (P,M,t)A\,\,\,\, . 
\end{equation}          

Then we find:

\begin{equation}   
  \frac{dn_D}{dt}=0,\,\,\,\,\,\,\,\, 
\frac{d <M>}{dt} = \frac{j_0}{n_D} \int d {\vec P} d M \sigma(M)
f_D(P,M,t) >0 \,\,\,\,\, ,
\end{equation}

\begin{eqnarray}
\frac{d U_{ab}(\lambda)}{dt}= &&-\frac{m j_0 (a+3b)}{3 M_0 n_D}
\int d {\vec P} d M \frac{P^a}{\lambda M^{b+1}} \sigma(M)f_D(P,M,t)
 \nonumber \\
&& + \frac{j_{\epsilon} a(a+1)} {3 M_0 n_D}
\int d {\vec P} d M \frac{P^{(a-2)}}{\lambda M^b} \sigma(M)f_D(P,M,t)\,\,\,.
\end{eqnarray}

 The question arises whether stationary averages in 
the limit $t \to \infty$ are possible. For example for the average 
kinetic energy of grains $E(t)=U_{21}(2)$ we find, as follows from 
Eq.9, the stationary solution in the limit $t \to \infty$ in the 
case $\sigma(M) \sim M$ $(\sigma\equiv \sigma_0^{\prime} M)$:
\begin{equation}
\lim_{t\to\infty} <E(t)>= \frac{6}{5} T_0 \,\,\, .
\end{equation}

We emphasize that if we formally omit the last term in 
Eq.4, describing change of the mass $M$, the stationary 
Maxwell distribution function for grains with the 
temperature $T_D=2T_0$ can be immediately found. This
result coincides with the solution obtained in [4] for
the limit of uncharged particles. Really the omitted term
is of the same order as other terms in Eq.4, as shown
above. Nevertheless the physical results and predictions 
obtained in [4] can be realized if the physical process 
of transfer of atoms from the surface of dust particles to 
the plasma take a place and is included in the kinetic 
theory. In this case the mass of dust particles can be 
fixed due to this process.
A more detailed analysis of the problem of stationary 
averages in the limit $t \to \infty$ for neutral systems and 
dusty plasmas in parallel with the consideration of the 
nonstationary solutions of Eq.4 for different cases,
in particular, of such as (for the case $\sigma(M)=$const):  
\begin{equation}
f_D(P,M,t) = \varphi(t-\frac{M}{m j_0 \sigma})\,\, \chi(P,M)
\end{equation}
and some others, including the solutions of generalized
Fokker-Planck equation for dusty plasmas, will be 
presented separately.  
\vskip 5mm
\begin{list}{{\bf Acknowledgments}} \item \end{list} 
The author would like to thank Drs. E.A.Allahyarov, 
W.Ebeling, A.M.Ignatov, G.M.W.Kroesen, S.A.Maiorov, 
P.P.J.M.Schram and A.G.Zagorodny for useful discussions 
on the kinetics of dusty plasmas.

\vskip 6mm

\begin{list}{{\bf References}} \item \end{list}
\hspace {-6.1mm}$[1]$ V.N.Tsytovich, O.Havnes, Comments Plasma Phys. Control 
    Fusion  15 (1995), p.267.\\ 
$[2]$ A.M.Ignatov, J.Physique IV, C4 (1997), p.215.\\
$[3]$ S.A.Trigger, P.P.J.M.Schram, J.Phys. D: Applied Phys. 32 (1999), p.234.\\
$[4]$ A.G.Zagorodny, P.P.J.M. Schram, S.A.Trigger, Phys.Rev.Lett. 84 (2000), p.3594.\\
$[5]$ P.P.J.M. Schram, A.G.Sitenko, S.A.Trigger, A.G. Zagorodny,
    Phys.Rev.E (to be published).\\ 
$[6]$ A.M.Ignatov, S.A.Maiorov, P.P.J.M.Schram, S.A.Trigger, 
    Short Communications in Physics, in print (Lebedev 
    Physical Institute, in Russian) (2000). \\
\vskip 0.3cm
S.A.Trigger, {\it email}:\,\,\,\, strig@gmx.net

\end{document}